\documentclass{article}
\usepackage[preprint]{spconf}
\copyrightnotice{\copyright\ IEEE 2019}
\toappear{To appear in {\it Proc.\ ASRU2019, December 14-18, 2019, Sentosa, Singapore}}
\usepackage{color,xcolor}
\usepackage{epsfig}
\usepackage{graphicx}

\usepackage{array}
\usepackage{booktabs}
\usepackage{colortbl}
\usepackage{multirow}
\usepackage{subcaption}
\usepackage{tablefootnote}

\usepackage{amsmath,amsfonts,amsthm,amssymb,bm}

\usepackage{changepage}
\usepackage{extramarks}
\usepackage{fancyhdr}
\usepackage{lastpage}
\usepackage{setspace}
\usepackage{soul}
\usepackage{xspace}

\usepackage{url}
\usepackage{hyperref}
\usepackage[numbers,sort&compress]{natbib}
\usepackage{algorithm, algorithmic}
\usepackage{enumitem}
\usepackage{verbatim} 
\newcommand{\sect}[1]{Section~\ref{sec:#1}}
\newcommand{\sectdot}[1]{Sec.~\ref{sec:#1}}
\newcommand{\ssect}[1]{Sec.~\ref{ssec:#1}}
\newcommand{\eqn}[1]{Eqn.~(\ref{eqn:#1})}
\newcommand{\fig}[1]{Fig.~\ref{fig:#1}}
\newcommand{\tbl}[1]{Table~\ref{tab:#1}}

\newcommand{\twosect}[2]{Sections~\ref{sec:#1} and \ref{sec:#2}}
\newcommand{\twoeqn}[2]{Eqns.~(\ref{eqn:#1}) and (\ref{eqn:#2})}

\newcommand{\twotbl}[2]{Tables~\ref{tab:#1} and \ref{tab:#2}}

\newcommand{\ignore}[1]{}

\DeclareMathOperator*{\argmax}{arg\,max}

\makeatletter
\DeclareRobustCommand\onedot{\futurelet\@let@token\@onedot}
\def\@onedot{\ifx\@let@token.\else.\null\fi\xspace}

\def\eg{\emph{e.g}\onedot} 
\def\ie{\emph{i.e}\onedot} 
\def\cf{\emph{c.f}\onedot} 
\def\etc{\emph{etc}\onedot} 
\def\wrt{w.r.t\onedot}

\makeatother

\definecolor{MyDarkBlue}{rgb}{0,0.08,1}
\definecolor{MyDarkGreen}{rgb}{0.02,0.6,0.02}
\definecolor{MyDarkRed}{rgb}{0.8,0.02,0.02}
\definecolor{MyDarkOrange}{rgb}{0.40,0.2,0.02}
\definecolor{MyPurple}{RGB}{111,0,255}
\definecolor{MyRed}{rgb}{1.0,0.0,0.0}
\definecolor{MyGold}{rgb}{0.75,0.6,0.12}
\definecolor{MyDarkgray}{rgb}{0.66, 0.66, 0.66}

\def\presec{\vspace{-0.4em}}
\def\postsec{\vspace{-0.4em}}

\def\W{\mathcal{W}}
\def\C{\mathcal{C}}
\def\O{\mathcal{O}}
\def\State{\mathcal{S}}

\title{Integrating Source-Channel and Attention-Based Sequence-to-Sequence Models for Speech Recognition}
\name{Qiujia Li, Chao Zhang \& Philip C. Woodland}
\address{Cambridge University Engineering Dept., Trumpington St., Cambridge, CB2 1PZ U.K.\\
\small{\texttt{\{ql264,cz277,pcw\}@eng.cam.ac.uk}}}

\begin{document}
\ninept

\maketitle
\begin{abstract}
This paper proposes a novel automatic speech recognition (ASR) framework called Integrated Source-Channel and Attention (ISCA) that combines the advantages of traditional systems based on the noisy source-channel model (SC) and end-to-end style systems using attention-based sequence-to-sequence models. The traditional SC system framework includes hidden Markov models and connectionist temporal classification (CTC) based acoustic models, language models (LMs), and a decoding procedure based on a lexicon, whereas the end-to-end style attention-based system jointly models the whole process with a single model. %By rescoring the hypotheses produced by SC-based models using an attention-based model, ISCA allows structured knowledge to be easily incorporated via the SC-based model while exploiting the complementarity of the attention-based model.
By rescoring the hypotheses produced by traditional systems using end-to-end style systems based on an extended noisy source-channel model, ISCA allows structured knowledge to be easily incorporated via the SC-based model while exploiting the complementarity of the attention-based model.
Experiments on the AMI meeting corpus show that ISCA is able to give a relative word error rate reduction up to 21\% over an individual system, and by 13\% over an alternative method which also involves combining CTC and attention-based models. 
%13\% relative for the multi-task trained system and 18\% relative for the separately trained system over the improved baseline.\ql{not sure what is the fair number here.}

%Currently, traditional hybrid models, connectionist temporal classification (CTC)-based models, and attention-based sequence-to-sequence models are three mainstream approaches to automatic speech recognition (ASR). In this paper, we treat hybrid and CTC models together as source-channel models in contrast to attention models based on their fundamental modelling methodology. The integrated source-channel and attention (ISCA) framework is proposed to combine these two types of systems. ISCA allows structured knowledge to be easily incorporated into the source-channel model while exploiting the complementarity of the attention model. Experiments on the AMI meeting corpus show that ISCA is able to improve the word error rate by 13\% relative for the multi-task trained system and 18\% relative for the separately trained system over the improved baseline.\ql{not sure what is the fair number here.}
\end{abstract}

\presec
\section{Introduction}
\postsec
\label{sec:intro}
% introduce SC models, lead to their drawbacks
Recently, automatic speech recognition (ASR) systems have obtained significant performance improvements due to the rapid development of deep learning~\cite{Hinton2012DeepNN}. Instead of traditional Gaussian mixture model (GMM)-based hidden Markov model (HMM) acoustic models and n-gram language models (LMs), state-of-the-art systems nowadays often use deep neural networks (DNNs) for acoustic and language modelling in the statistical ASR framework based on the modular \emph{noisy source-channel model} (SC)~\cite{jelinek1997statistical}. Along with the statistical models, an SC-based system also incorporates acoustic, phonetic, and lexical knowledge \etc via feature extraction, subword unit construction and decoding, which have demonstrated their significance in improving ASR accuracy and robustness over the years. Despite its good performance, building SC-based ASR systems is considered to be complex and usually requires additional resources such as a phonetic lexicon created by experts.

% introduce attention models (this is too long, move to later section)
% As an alternative to the modularised SC system, the attention-based encoder-decoder networks address ASR from a sequence-to-sequence modelling perspective. Originated from neural machine translation, recurrent neural networks (RNNs) are able to encode a variable-length input sequence from a source language into intermediate hidden representations, and then decode to obtain the output sequence in a target language~\cite{Cho2014LearningPR,Sutskever2014SequenceTS}. The introduction of the attention mechanism further improves the performance on machine translation tasks by solving the problem of information bottleneck caused by the fixed-length hidden representation~\cite{Bahdanau2015NeuralMT}. The term \emph{attention model} is used in the rest of the paper to refer to the attention-based sequence-to-sequence or encoder-decoder models.

% attention models in ASR, integration of knowledge
The attention-based encoder-decoder or sequence-to-sequence networks (referred to as attention-based models below) provide a promising alternative to the SC-based models. Shortly after their success in machine translation~\cite{Bahdanau2015NeuralMT}, attention-based models have been applied to ASR~\cite{Chorowski2015AttentionBasedMF,Lu2015ASO,Bahdanau2016EndtoendAL,Chan2016ListenAA}. This approach is considered to be \emph{end-to-end} because the attention-based model serves as an integrated acoustic and language model that does not require a lexicon when modelling with grapheme-based units. As the attention-based end-to-end approach greatly simplifies construction of an ASR pipeline, various ASR-specific techniques have been subsequently proposed to improve performance. Different encoder architectures have been explored to better handle the input acoustic features~\cite{Chan2016ListenAA,zhang2016EndtoEndSpeechRecognition,zhang2017VeryDeepConvolutional,dong2018SpeechTransformerNorecurrenceSequencetosequence,pham2019VeryDeepSelfAttention}. Improvements to the loss functions~\cite{prabhavalkar2018MinimumWordError,sabour2019OPTIMALCOMPLETIONDISTILLATION,cui2018ImprovingAttentionBasedEndtoEnd}, data augmentation~\cite{hayashi2018BackTranslationStyleDataAugmentation,park2019SpecAugmentSimpleData}, and other optimisation related methods~\cite{karita2018SequenceTrainingEncoderdecoder,tjandra2018SequencetoSequenceASROptimization,zhou2018ImprovingEndtoEndSpeech,lu2016TrainingRecurrentNeural} have been found to be effective. Meanwhile, some speaker-related knowledge has been integrated into the end-to-end system~\cite{delcroix2018AuxiliaryFeatureBased,ochiai2018speaker}. Alternative subword units have been studied to incorporate richer information~\cite{zweig2017AdvancesAllneuralSpeech,chan2017LatentSequenceDecompositions,xu2019ImprovingEndtoendSpeech}. %Prefix-tree based decoding algorithm facilitates the use of a word-level LM based on a vocabulary \cite{**}, which enables the 
Recurrent neural network language models (RNNLMs) trained on additional text data have been integrated into the attention-based systems during training or testing~\cite{hori2018EndtoendSpeechRecognition,toshniwal2018ComparisonTechniquesLanguage,shan2019ComponentFusionLearning}. %The integration of additional vocabulary and LMs allow the use of extra linguistic information and text-only training data
Together, despite having a large numbers of parameters and requiring a considerable amount of training data, recent attention-based systems have achieved state-of-the-art results on some ASR tasks~\cite{chiu2018StateoftheartSpeechRecognition,park2019SpecAugmentSimpleData,luscher2019RWTHASRSystems}, which makes the end-to-end approach increasingly competitive and encouraging.

% Such an encoder-decoder and attention-based DNN structure is named as the \emph{sequence-to-sequence model} \cite{**}, which is sometime also referred to as the \emph{attention-based framework} to stress the importance of attention \cite{**}. 
%Since the sequence-to-sequence model often works by mapping a variable-length input sequence into some embedding vectors with a DNN (encoder), and then mapping the vectors to the output sequence using another DNN (decoder) and attention models, it is also referred to as encoder-decoder method or attention-based method. 
%Since the sequence-to-sequence model often relies on the ``attention'' mechanism to map the input sequence to the output one, it is also referred to as an attention-based method.
%Often based on the ``attention'' mechanism \cite{**},  sequence-to-sequence model first encodes a variable-length input sequence into some intermediate representations using its encoder part, and then decodes them to produce the output sequence using its decoder part. The encoder, decoder, and attention parts are all formed by DNN structures

% introduce CTC, both e2e and SC
Besides the attention-based systems, end-to-end ASR approaches also include connectionist temporal classification (CTC)-based methods~\cite{Graves2006CTC,Graves2012SequenceTW}. CTC is a sequence-level objective function which does not require frame-level alignments produced by an existing ASR system for DNN acoustic model training. Hence with context-independent graphemic output units, these systems are considered to be end-to-end as lexicons and context-dependent subword unit clustering are not needed. Since CTC is almost mathematically equivalent to HMM-based models~\cite{Zeyer2017CTCIT,Hadian2018EndtoendSR}, this paper treats CTC-based models as one special class of SC-based systems. %A variant of CTC model is the RNN-transducer, which has a prediction network serving as the LM that is jointly trained with the CTC acoustic model, and is therefore also end-to-end \cite{**}.

% RNN transducer~\cite{Graves2012SequenceTW} is a variant of CTC-based models where a prediction network is trained jointly with the CTC transcription model, which can be viewed as a joint training framework of acoustic and language models.
% Moreover, word units can be directly used for acoustic modelling in CTC that can remove the dictionary as well \cite{**}.
%By extending SC with an additional \textit{spell model}, an attention-based end-to-end system can be used to rescore each hypothesised word sequence in the (speech) frame-synchronised decoding procedure of the SC system. 

In this paper, a framework, Integrated Source-Channel and Attention (ISCA), is proposed to combine the advantages from both the SC-based models and the attention-based models for ASR. By extending the noisy source-channel model that covers both DNN-HMMs and CTC models, the attention-based model can be used as an additional complementary module to rescore each hypothesised word sequence from the (speech) frame-synchronous decoding procedure of the SC-based system. This enables the joint use of attention models with phonetic and linguistic knowledge in a simple way.
Inspired by~\cite{kim2017JointCTCattentionBased,Watanabe2017HybridCA}, the hidden layers of the acoustic model in the SC-based system and the encoder of the attention-based system can be shared via \emph{multi-task learning}. Alternatively, the two systems can be constructed independently, which makes ISCA a general approach that combines a traditional ASR system with an attention-based end-to-end system. First, commonly used setups of both the attention-based model and the CTC model baselines are improved by drawing techniques from DNN-HMM acoustic models. Then, experimental results using the augmented multi-party interaction (AMI) dataset show that ISCA improves the word error rates (WERs) over both component systems by significant margins. 

% If the SC system is a CTC based end-to-end system, ISCA is end-to-end since it combines two end-to-end systems with additional information. Otherwise, the attention-based model can be viewed as an extra graphemic unit based LM trained with paralleled acoustic and text data. 

This paper is organised as follows. \twosect{scmodel}{att} describe the SC and attention-based frameworks. \sect{isca} presents the details of the ISCA framework. Experimental setup and results are given in \sectdot{setup} and \sectdot{exp} with conclusions in \sectdot{conclusion}.

\presec
\section{Noisy Source-Channel Model}
\postsec
\label{sec:scmodel}
Speech recognition can be viewed as a \emph{noisy source-channel} modelling problem~\cite{jelinek1997statistical} where speech is produced and encoded via a noisy channel and the recogniser is to find the most probable source text $\W^{*}$ given the output of the noisy channel $\O$. According to Bayes' rule, decoding follows the \emph{maximum a posteriori} (MAP) rule to search over each possible hypothesis $\W$ by
\begin{equation}
    P(\W|\O) \propto p(\O|\W)P(\W),
    \label{eqn:scmodel}
\end{equation}
% \begin{equation}
%     P(\bm{w}^{\text{ref}}|\bm{o}) \propto \max_{\bm{w}}\,p(\bm{o}|\bm{w})P(\bm{w})
%     \label{eqn:scmodel}
% \end{equation}
where $p(\O|\W)$, estimated by an acoustic model, is the likelihood of generating the observation through the channel; $P(\W)$, approximated by an LM, describes the underlying probabilistic distribution of the source. In this way, an SC-based ASR system consists of several independent modules. 

\presec
\subsection{HMM Acoustic Models}
\postsec
\label{ssec:hmm}
In acoustic modelling, HMMs, together with GMMs or DNNs, are used to model the generative process of the observation sequences with respect to the subword units. The parameters of the HMMs are estimated using the maximum likelihood (ML) criterion:
\begin{equation}
    %\mathcal{F}_{\text{ML}} = \sum_{u} \log \sum_{\bm{s}^{(u)}}\prod_{t=1}^{T_u} P(s_{t+1}|s_t)p(\bm{o}_t^{(u)}|s_t)
    \mathcal{F}_{\text{ML}} = \ln \sum_{\State}\prod_{t=1}^{T} P(s_{t+1}|s_t)p(\bm{o}_t|s_t),
    \label{eqn:hmm}
\end{equation}
where $[s_{1},s_{2},\ldots,s_{T}]$ is a possible HMM state sequence for the given transcription $\W^\text{ref}$ of an utterance, whose observation $\O$ is $[\bm{o}_1,\bm{o}_2,\dots,\bm{o}_T]$. The observation probability $p(\bm{o}_t|s_t)$ is estimated by a GMM or a DNN, and $P(s_{t+1}|s_t)$ is the transition probability.
%where $\bm{s}^{(u)}$ is all possible HMM state sequences for a given word sequence $\bm{w}^{(u)}$ of the corresponding utterance $u$. An observation probability $p(\bm{o}_t^{(u)}|s_t)$ can be estimated using a GMM or a DNN.

%based on the \emph{forward-backward procedure} \cite{**}. 
The GMM-HMM parameters are often trained at the sequence-level by using the \emph{forward-backward procedure} to find the alignments between time steps and all possible state sequences $\State$ corresponding to $\W^{\text{ref}}$~\cite{Dahl2012ContextDependentPD,Hinton2012DeepNN}. 
%The GMM-HMM parameters are often trained at sequence-level by first joining multiple HMMs into a composite HMM related to $\bm{w}$ and performing \emph{forward-backward procedure} to find the alignments between all possible $\bm{s}$ and the time steps \cite{**}. 
%based on the \emph{forward-backward procedure} \cite{**}. 
%At training time, the GMM-HMM parameters can be computed efficiently using 
For DNN-HMMs, the DNN is often trained as a classifier to estimate the state posterior probability $P(s_t|\bm{o}_t)$, whose training alignments are generated by a pre-trained system.
%For DNN-HMMs, the DNN is often trained as a classifier to estimate the state posterior probability $P(s_t|\bm{o}_t)$, whose training alignments generated by an existing system are used to provide the frame-level target distributions to learn using the cross-entropy (CE) criterion. 
Such frame-level DNN training allows better data shuffling which was found to be important for efficient training with stochastic gradient descent (SGD)~\cite{Su2013ErrorBP}.
The log-likelihood $\log p(\bm{o}_t|s_t)$ is calculated according to Bayes' rule by
\begin{equation}
     \ln p(\bm{o}_t|s_t) = \ln P(s_t|\bm{o}_t) - \ln P(s_t) + \ln p(\bm{o}_t),
    \label{eqn:prior}
\end{equation}
where $P(s_t)$ is the prior probability estimated by the frequency of $s_t$ from the training alignments and $p(\bm{o}_t)$ is the observation prior. Alternatively, the forward-backward procedure can also be applied to train DNN-HMMs at the sequence-level~\cite{Bridle1990AlphanetsAR,seniorThesis}. %alpha-net, bengio's thesis, volchev's thesis, kingsbry's paper etc.
%and the HMM transition probabilities $P(s_{t+1}|s_t)$ are not trained since 
%To obtain the likelihood $p(\bm{o}_t|s_t)$, the posterior probability from the DNN needs to be divided by the training data prior $P(s_t)$ calculated from the number of frames aligned to each output target.
% \begin{equation}
%     p(\bm{o}|s_i)\propto \dfrac{P(s_i|\bm{o})}{P(s_i)}
% \end{equation}

Since phones often serve as natural units to define word pronunciations, phonetic-based units are commonly used in conjunction with a lexicon. %Although some linguistic expertise is required, lexicons for major languages are generally available.
Although the \emph{first-order Markovian property} of HMMs assumes the transition to each HMM state only depends on its immediate preceding one, the effect of the assumption can be reduced by introducing more contextual information, \eg DNN models taking the current and adjacent frames as input~\cite{peddinti2015time,Kreyssig2018ImprovedTU}, and context-dependent phones clustered by phonetic decision trees~\cite{Young1994TreeBasedST}.
%by technologies like context-dependent (CD) modelling, which akin to the use of HMMs to model the CD phones and cluster them into fewer classes with phonetic decision trees \cite{**}.
% It is worth noting that Graphemic subword units can be used for HMM-based acoustic models as well \cite{**}. 
Minimum error rate discriminative sequence training is also useful for acoustic models to learn information beyond the scope of individual frames~\cite{Woodland2002LargeSD,Povey2002MinimumPE}. LMs provide word-level contextual information estimated using a text corpus where supplementary text data is often available in addition to speech transcriptions. Many state-of-the-art ASR systems are SC-based that combines information from all the aforementioned sources~\cite{xiong2017toward}.

% and enlarged feature windows. Furthermore, complex neural network architectures allow time and frequency information to be further exploited. SC decoders can also incorporate high-order n-gram language models during the first-pass. Since CE-trained acoustic models are optimised for frame-level classification accuracy, rather than the final word error rate (WER), discriminative sequence training are designed to alleviate the criterion mismatch for acoustic modelling. By cascading all the above techniques and many more, HMM-based models are currently dominant in deployed state-of-the-art ASR systems.

% At time $t$, the likelihood of generating the observation $\bm{x}_t$ is
% \begin{equation}
%     p(\bm{x}_t|s) = \dfrac{P(s|\bm{x}_t)p(\bm{x}_t)}{P(s)}
% \end{equation}
% where $\bm{x}_t$ is the input feature to the neural network or the noisy observation, $s$ is the HMM state.
\presec
\subsection{CTC Acoustic Model}
\postsec
\label{ssec:ctc}
CTC is a method that trains a DNN acoustic model with a blank output symbol at the sequence-level without any explicit HMM structure~\cite{Graves2006CTC}. The training objective function is
%Another form of source-channel models uses connectionist temporal classification (CTC) where the sequence-level objective function is to maximise the likelihoods of target labels with all possible alignments~\cite{Graves2006CTC}. The blank unit allows probabilities of all possible alignments to be computed using the forward-backward algorithm.
\begin{align}
    %\mathcal{F}_{\text{CTC}} = \sum_{u}\log \sum_{\bm{\pi}^{(u)}}\prod_{t=1}^{T_u}P(\pi_t|\bm{o}_t^{(u)})
    \mathcal{F}_{\text{CTC}} = \ln \sum_{\State}\prod_{t=1}^{T}P(s_t|\bm{o}_t),
    \label{eqn:ctc}
\end{align}
where $\State$ are all possible symbol sequences that can map to $\W^\text{ref}$ by removing repeated symbols and the blanks. During training, alignments and losses are computed by the forward-backward procedure.  
%where $\State = \{\State|\mathcal{B}(\State) = \W^\text{ref}\}$ and $\mathcal{B}(\cdot)$ is a many-to-one mapping that removes repeated symbols and the blanks. During training, all possible alignments $\State'$ are computed by forward-backward procedure, which is the same as HMMs.  

%where $\bm{\pi^{(u)}} = \{\bm{\pi}|\mathcal{B}(\bm{\pi}) = \bm{w}^{(u)}\}$ and $\mathcal{B}(\cdot)$ is a many-to-one mapping that removes repeated symbols and the blanks.
\begin{figure}[ht]
    \centering
        \includegraphics[width=.49\linewidth]{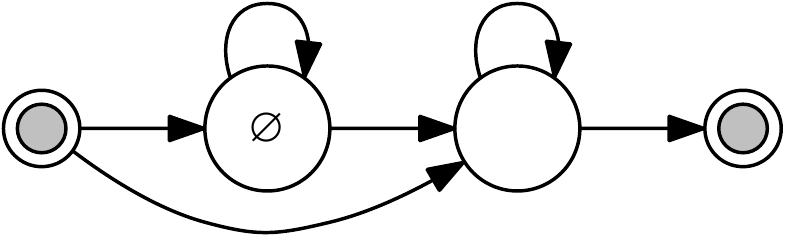}
        %\caption{A CTC equivalent HMM topology}
    \caption{A CTC-equivalent HMM topology. The white and grey cycles are emission and non-emission states. $\varnothing$ is the blank symbol.}
    \label{fig:hmm&ctc}
\end{figure}

%\begin{figure}[t]
%    \centering
%    \begin{subfigure}{.49\linewidth}
%        \centering
%        \includegraphics[width=0.7\linewidth]{fig/HMM.pdf}
%        \vspace{0.5em}
%        \caption{Single-state HMM topology}
%    \end{subfigure}
%    \begin{subfigure}{.49\linewidth}
%        \centering
%        \includegraphics[width=\linewidth]{fig/CTC.pdf}
%        \caption{A CTC equivalent HMM topology}
%    \end{subfigure}
%    \caption{A relationship between HMM and CTC. The white and grey cycles are emission and non-emission HMM states, and $\varnothing$ stands for the CTC blank symbol.}
%    \label{fig:hmm&ctc}
%\end{figure}
%By comparing \eqn{ctc} to \twoeqn{hmm}{prior},CTC is equivalent to a special instantiation of the 2-state HMM structure when state priors $P(s_t)$, observation priors $p(\bm{o}_t)$, and transition probabilities ${P(s_{t+1}|s_t)}$ are equal~\cite{Hadian2018EndtoendSR}. 
By comparing \eqn{ctc} to \twoeqn{hmm}{prior}, CTC is equivalent to a special instantiation of the 2-state HMM structure when $P(s_t)$, $p(\bm{o}_t)$, and ${P(s_{t+1}|s_t)}$ are ignored~\cite{Hadian2018EndtoendSR}. 
As shown in \fig{hmm&ctc}, the first emission state of the HMM is the skippable blank state ($\varnothing$) with a self-loop and the second one corresponds to the subword unit. The blank state is shared across all HMMs. %It has been shown for DNN-HMMs, ignoring the non-zero HMM transition probabilities make almost no differences in practice. In \sect{exp}, the effect of state priors of CTC will be demonstrated experimentally. Therefore, by modifying the HMM structure accordingly, CTC-based models are a type of acoustic model in an SC-based system. The RNN-transducer is a variant of CTC-based models where a prediction network is jointly trained and acts as a language model in parallel with the acoustic model. Compared to HMM-based models, CTC-based models do not need forced alignments by a prior system. As a result, frame-level shuffling during the training of a CTC-based model is infeasible.

In practice, $p(\bm{o}_t)$ is a constant and ${P(s_{t+1}|s_t)}$ makes little difference when being forced to set to 1.0~\cite{Dahl2012ContextDependentPD}. At test time, the posterior probability of the CTC blank symbol can be penalised by an extra empirical value~\cite{sak2015FastAccurateRecurrent,sak2015LearningAcousticFramea}, which can be seen as a rough approximation of the prior $P(s_t)$. %Therefore, CTC can be basically viewed as a method that combines the HMM topology in \fig{hmm&ctc}, ML training criterion and the forward-backward procedure, which makes it an acoustic model training method in the SC framework.
As a result, it is reasonable to view CTC as an acoustic modelling method in the SC framework, which combines the HMM topology in \fig{hmm&ctc}, the ML training criterion and the forward-backward procedure. In addition, the RNN-transducer~\cite{Graves2012SequenceTW} can be viewed as an SC-based system, since it extends CTC by introducing an extra prediction network as a language model trained in parallel with the CTC acoustic model.% to replace LM and decoder. 

\presec
\section{Attention-based Model}
\label{sec:att}
\subsection{Attention-based Sequence-to-sequence Model}
\postsec
\label{ssec:seq2seq}
%Originating from neural machine translation, recurrent neural networks (RNNs) are able to encode a variable-length input sequence from a source language into intermediate hidden representations, and then decode to obtain the output sequence in a target language~\cite{Cho2014LearningPR,Sutskever2014SequenceTS}. The attention mechanism further improves the performance on machine translation tasks by solving the problem of information bottleneck caused by the fixed-length hidden representation~\cite{Bahdanau2015NeuralMT}. Compared to machine translation, speech recognition is also a sequence-to-sequence problem and the attention models provide a solution alternative to the SC framework.
%, although the input-to-output relationship is strictly monotonic and many-to-one. Mathematically, attention models provide an alternative view of addresses the speech recognition problem.
Originating from machine translation, the attention-based sequence-to-sequence model maps from the input sequence of a source language to the output sequence of a target language~\cite{Bahdanau2015NeuralMT}.
For ASR, the attention model maps a $T$-length input observation sequence $\O$ to an $I$-length output subword unit sequence $\C$.
Instead of decomposing into an acoustic model and a language model as in \sectdot{scmodel}, attention-based models compute the posterior distribution $P(\C|\O)$ directly following the \emph{chain rule of conditional probability}:
\begin{equation}
    P(\C|\O) = P(c_1|\O)P(c_2|c_1,\O)\dots P(c_I|c_1,\dots,c_{I-1},\O),
    \label{eqn:seqprob}
\end{equation}
where $\C$ is converted from $\W^\text{ref}$ for training. Compared to \eqn{scmodel}, the attention-based systems hold some theoretical advantages over the SC-based systems: they do not necessarily rely on the first-order Markovian assumption during decoding, and the acoustic and language information is jointly learned using a single model without making any independence assumptions.

The attention-based model consists of a neural encoder, a neural decoder, and an attention mechanism. The neural encoder maps the variable-length input sequence $\mathcal{O}$ to an intermediate embedding $\mathcal{E}$. At each decoding step $i$, the encoded information $\mathcal{E}$ is first transformed into a context vector $\bm{h}_{i}$ based on the annotation vector $\bm{a}_{i}$ produced by the attention mechanism. Then, the neural decoder transforms the hidden information $\bm{h}_{i}$ to the posterior distribution on the current subword unit $c_i$ based on the previous output and the previous decoder state $\bm{d}_{i-1}$. The procedure stops when the \textit{end-of-sentence} symbol is decoded, which allows output sequences to have variable lengths. More specifically,
%\begin{align*}
%\bm{e}&=\text{neural encoder}(\bm{o})\\
%\bm{h}_n&=\text{attention mechanism}(c_{1},c_{2},\ldots,c_{n-1},\bm{e})\\
%c_n&=\text{neural decoder}(c_{1},c_{2},\ldots,c_{n-1},\bm{h}_n).
%\end{align*}
\begin{align}
\mathcal{E} & = \text{NeuralEncoder}(\mathcal{O})\\
\bm{a}_i & = \text{AttentionMechanism}(\bm{a}_{i-1},\bm{d}_{i-1},\mathcal{E})\\
\bm{h}_i & = \mathcal{E}\;\bm{a}_{i}\\
% \bm{h}_n&=\bm{a}_n\;[\bm{e}_1,\bm{e}_2,\ldots,\bm{e}_T]^{\text T}\\
c_i,\bm{d}_{i} & = \text{NeuralDecoder}(c_{i-1},\bm{d}_{i-1},\bm{h}_i).\label{eqn:neuraldecoder}
\end{align}

Despite the elegance of this framework, there are some differences when applying it to ASR. For machine translation, the attention mechanism is particularly effective because the input and output sequences generally have similar lengths and irregular alignments. For ASR, the input observation sequence is often much longer than the output subword unit sequence, and their alignment is relatively local and strictly monotonic~\cite{zhang2019WindowedAttentionMechanisms}. Thus the strength of the attention mechanism may not be brought fully into play, while making it challenging to process streaming data~\cite{Chiu2018MonotonicCA}. Furthermore, the attention-based systems often require more training data and model parameters than the SC-based systems~\cite{park2019SpecAugmentSimpleData,luscher2019RWTHASRSystems} since it is not straightforward to incorporate rule-based knowledge such as a lexicon~\cite{sainath2018NoNeedLexicon}. 
% An LM in SC-based system can also be estimated from a large amount of text-only data, whereas the audio and text training data need to be paired for the attention-based system due to the integration of acoustic and language models. 
Moreover, since the previous decoding output needs to be fed into the neural decoder to obtain the probability distribution over the next subword unit, it is very expensive to search over a large hypothesis space and store rich decoding results in the form of a lattice. 
%To overcome these issues, various changes have been made to the attention mechanism \cite{**}, choice of subword units \cite{zweig2017AdvancesAllneuralSpeech,chan2017LatentSequenceDecompositions,xu2019ImprovingEndtoendSpeech}, decoding algorithm \cite{**}, and the integration of RNNLMs \cite{**}.
As described in \sectdot{intro}, various changes have been made to the attention-based end-to-end framework to address some of the issues above. 
% To overcome these issues, various ASR-specific changes have been made to the attention-based end-to-end framework~\cite{}.

%In order to overcome the aforementioned issues, the attention mechanism was modified to generate monotonic and more local alignments \cite{**}. Different grapheme-based subword units were studied to provide richer information without a lexicon.  

%require a significant amount of data and model with a considerable size of model to learn the highly complex mapping between acoustic signal and text as it cannot easily incorporate structured data and other rule-based knowledge. Experiments have shown that using another RNNLM, either trained jointly or separately, can also improve the performance of the system significantly. For the attention model to be decoded without a lexicon, grapheme-based units are normally used. Since the output of the attention model is frame-asynchronous and the current output needs to be fed into the decoder to obtain the probability distribution over the next symbol, the decoding result cannot be stored in forms of lattices or confusion networks as for SC models. 

\presec
\subsection{Multi-task Training of Hybrid CTC and Attention Models}
\postsec
\label{ssec:multitask}
Within the scope of end-to-end speech recognition, a hybrid CTC and attention architecture has been proposed to take advantage of both CTC and attention models during both training and decoding~\cite{Watanabe2017HybridCA}. This architecture adopts the multi-task training method with one output branch using the CTC loss and the other branch using the loss of the attention-based decoder. The rest of the acoustic model and the neural encoder share the same parameters. During decoding, a beam search is performed on the attention branch, where the prefix score of the current partial hypothesis is obtained from the CTC branch. The CTC prefix score and the attention score are then interpolated for ranking and pruning during the search~\cite{Watanabe2017HybridCA}. A score from a multi-level RNNLM or a look-ahead word-based RNNLM~\cite{hori2018EndtoendSpeechRecognition} can also be added to the joint score during the beam search. The benefits of this architecture are that each branch provides a certain degree of regularisation for the other branch and the score interpolation during decoding refines the search space. However, this framework requires both branches to use identical subword units as well as the same encoder parameters, which reduces their complementarity to some extent. Optimising the interpolation coefficients for both training and decoding is very important for the performance, which requires the model to be re-trained and the dev set to be decoded multiple times to achieve near-optimal results. %It is also hard to process streaming data.
\presec
\section{Integrated Source-Channel and Attention}
\postsec
\label{sec:isca}
\begin{figure}[t]
    \centering
    \includegraphics[width=\linewidth]{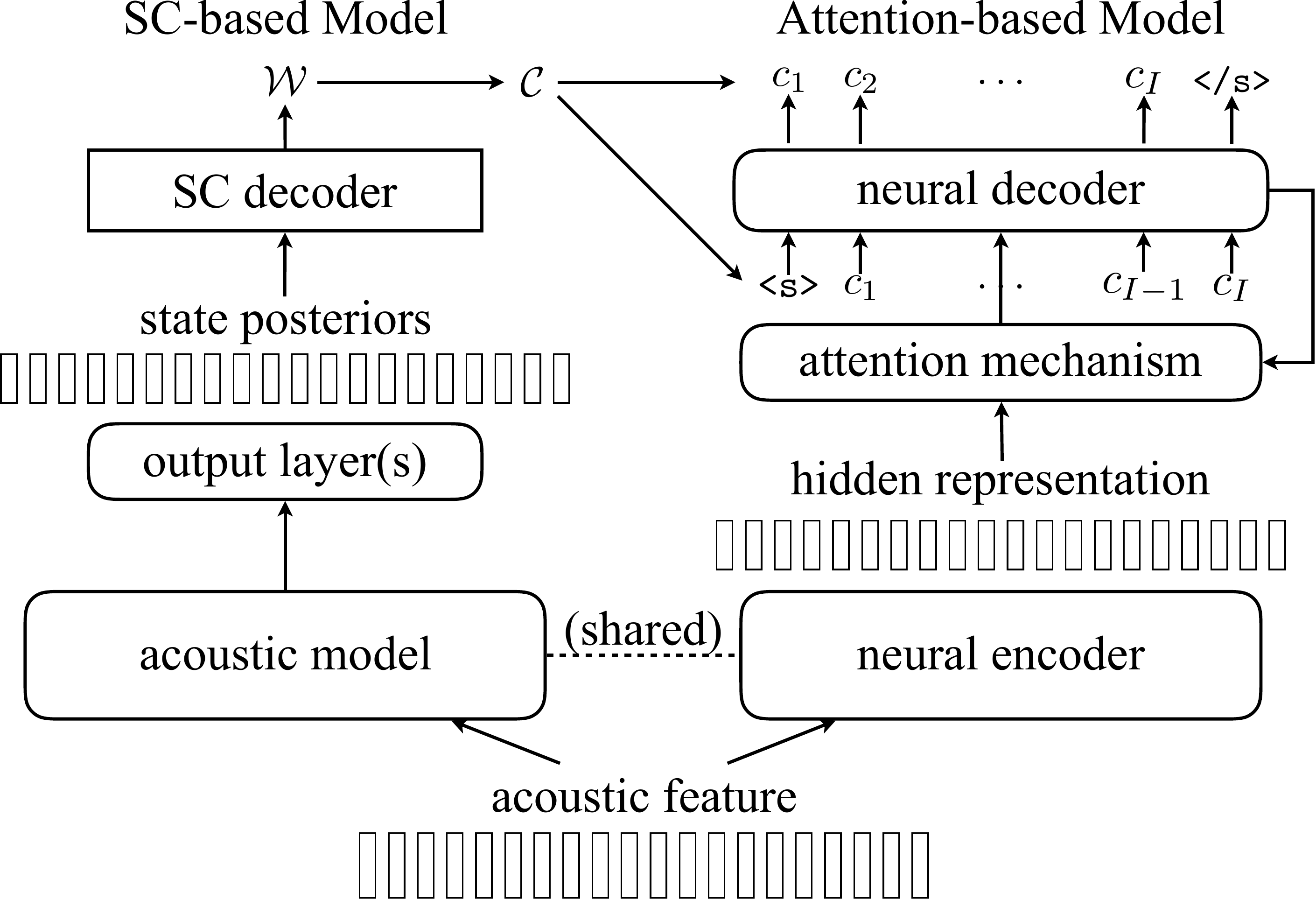}
    \caption{Integrated source-channel and attention (ISCA) framework. Modules in rounded boxes are trainable.}
    \label{fig:isca}
    \vspace{-2em}
\end{figure}
The proposed ISCA framework that combines both an SC-based model and an attention-based model is described in this section. Compared to the hybrid CTC and attention architecture reviewed in \ssect{multitask}, ISCA %by reversing the functions of the two systems during decoding,
performs frame-synchronous decoding using the SC-based model first and then rescores hypotheses with the attention-based model. %This makes it not only easy to incorporate extra phonetic and linguistic knowledge, but also capable of handling streaming service with a second pass. 
This integration of an SC-based model means ISCA both easily incorporates extra phonetic and linguistic knowledge, and is capable of handling streaming data in the first decoding pass. %Furthermore, it is worth noting that apart from the traditional frame-synchronised style, the SC decoder can also be an RNN-based decoder, if RNN-transducer can be viewed as an extended SC system.

%To combine multiple SC models, standard methods such as confusion network combination and ROVER work better when based on decoding lattices. However, it is not straightforward to generate lattices using attention models, cannot generate lattices due to its autoregressive nature. and computing their hypothesis-level confidence scores requires further deliberation. Therefore, both traditional model combination approaches are no longer applicable. Since the SC model and the attention model are very dissimilar in terms of the underlying distribution modelled, the training objective, and the decoding procedure. This section introduces the ISCA framework to explore the complementarity between these two classes of models.

\postsec\postsec
\subsection{Framework Overview}
\postsec
In the system described in \ssect{multitask}, the decoding process is based on the attention-based model where the CTC branch plays an auxiliary role. Owing to the advantages of SC-based models, especially the relative ease of integrating structured knowledge into the pipeline, the ISCA framework focuses on the SC-based system and then establishes the dependence between the SC-based model and the attention-based model. Given an SC-based model that maximises the posterior probabilities of a word sequence $\W$ with an acoustic model $p(\O|\W)$ and a language model $P(\W)$, ISCA is an \textit{extended noisy source-channel model} that includes an extra attention-based model trained to maximise the probability of the corresponding correct subword unit sequence $\C$. The posterior probabilities given by the extended SC can be obtained as follows:
\begin{align}
    P(\W|\O) & = \sum_{\C} P(\W, \C|\O)\nonumber\\
    & = \sum_{\C} P(\C|\W, \O) P(\W|\O)\nonumber\\
    & \propto p(\O|\W)P(\W)\sum_{\C} P(\C|\W,\O).
    \label{eqn:isca}
    \vspace{-1em}
%    P(\W|\O) & = \sum_{\C} P(\W, \C|\O) = \sum_{\C} P(\C|\W, \O) P(\W|\O)\nonumber\\
%    % & \nonumber\\
%    & \propto p(\O|\W)P(\W)\sum_{\C\in\mathcal{L}(\W)} P(\C|\O)
%    \label{eqn:isca}
\end{align}
%From \eqn{seqprob}, an attention model directly computes $P(\C|\O)$. It can be conditioned on a specified word hypothesis $\W$ by rescoring 

Since an attention-based model directly computes $P(\C|\O)$ from \eqn{seqprob}, it can be further conditioned on $\W$ to compute $P(\C|\W,\O)$ by rescoring $\W$ with the attention model. If $\C$ is a subword unit sequence obtained by converting $\W$ using a lexicon, there is $P(\C|\W,\O)=P(\C|\O)$; otherwise $P(\C|\W,\O)=0$.
If the subword units are graphemic, then there is often only one $\C$ for each $\W$. If the subword units are phonetic, there are often multiple $\C$s due to multiple pronunciations, and 
 $\sum_{\C} P(\C|\O)$ is the sum of posteriors of all possible phonetic sequences for a given $\W$. 
%where $\mathcal{L}$ is a mapping function from the word sequence $\W$ to the sequence of subword units output by the attention model, \ie a graphemic or phonetic lexicon. If the subword units are graphemic, then $\mathcal{L}$ is a one-to-one decomposition and the summation in \eqn{isca} becomes redundant. If the subword units are phonetic, converting a word sequence to a phonetic sequence can be a one-to-many mapping due to heteronyms. $\sum_{\C\in\mathcal{L}(\W)} P(\C|\O)$ is the sum of posteriors of all possible phonetic sequences for a given word hypothesis $\W$. 
%from the source-channel model and the corresponding input acoustic features.
The attention-based model can also be viewed as a special language model conditioned on acoustics. 

In contrast to ISCA, standard combination methods, such as confusion network combination~\cite{Evermann2000PosteriorPD} and ROVER~\cite{Fiscus1997APS}, are not suitable for attention-based models, since they normally require decoding lattices or confidence scores from both systems, in order to work well. % Specifically, for the attention-based model, it is often overly computational expensive to generate decoding lattices.

%complexity for forwarding the attention model to generate decoding lattice is exponential. 

%To combine multiple SC models, standard methods such as confusion network combination and ROVER work better when based on decoding lattices. However, it is not straightforward to generate lattices using attention models, cannot generate lattices due to its autoregressive nature. and computing their hypothesis-level confidence scores requires further deliberation. Therefore, both traditional model combination approaches are no longer applicable. Since the SC model and the attention model are very dissimilar in terms of the underlying distribution modelled, the training objective, and the decoding procedure. This section introduces the ISCA framework to explore the complementarity between these two classes of models.

%based either on graphemes or phones.

\presec
\subsection{Subword Unit Selection}
\postsec
% Lexicons are very important for acoustic modelling and different subword units correspond to different lexicons as in \tbl{lexicon}. For the source-channel model, a lexicon can be added to the decoding process to constrain the search space such that only paths in the lexicon are allowed. 
% \begin{table}[h]
%     \centering
%     \begin{tabular}{c|ll}
%         subword unit & example \\
%         \midrule
%         grapheme & \texttt{h\;e\;l\;l\;o} & \\
%         % PEG & \texttt{h$^\text{I}$\;e$^\text{M}$\;l$^\text{M}$\;l$^\text{M}$\;o$^\text{F}$} & \\
%         phone & \texttt{hh\;ax\;l\;ow} & \texttt{hh\;eh\;l\;ow} \\
%     \end{tabular}
%     \caption{Sequences of different subword units for ``hello''.}
%     \label{tab:lexicon}
% \end{table}
Although graphemic lexicons are very easy to construct, phonetic lexicons often help reduce the acoustic modelling complexity, especially for languages like English with irregular orthography. Context-dependent modelling can be applied to both graphemes and phones, and due to a large number of possible context-dependent units, decision-tree-based clustering is often needed to reduce the number of classes \cite{Young1994TreeBasedST}. %Decoding process also need to take adjacent context into consideration. 

Compared to the multi-task trained CTC and attention-based framework where the subword units from both branches have to be identical, the ISCA framework allows the SC-based model and the attention-based system to have different modelling units, either graphemic or phonetic, context-independent or -dependent, and the scores from the two models are integrated at the word-level regardless of their respective lexicons.
%In the ISCA framework, the source channel model can have one lexicon while the attention model can have another. 
It is predictable that for SC-based models, context-dependent modelling is more significant since it reduces the effect of the first-order Markovian assumption by conditioning on the preceding and/or succeeding subword units. For the attention-based models, the effectiveness of using context-dependent units is questionable as it has already conditioned on all previous subword units from \eqn{seqprob}.

%of the lexicon is greater as the underlying HMM structure imposes a conditional independence assumption and the use of context-dependent units with decision trees have been shown to improve the performance of acoustic models. The lexicon is normally integrated into the search graph in either weighted finite state transducer decoder or lexical-tree-based decoder. However, for attention models, the effectiveness of using context-dependent units is rather questionable. For every step in the attention-based decoder, the result from the previous stage is directly fed into the next stage as an input.
\presec
\subsection{Training}
\postsec
As shown in \fig{isca}, the SC-based model and the attention-based model can be trained separately or together in a multi-task fashion by sharing the neural encoder parameters with the hidden layers of the acoustic model. %where the parameters neural encoder of the attention model are shared.
For multi-task trained models, the total number of parameters in the entire system may be much smaller due to parameter sharing. Although multi-task training can be an effective type of regularisation, setting the interpolation weights between the two losses and configuring the learning rate to achieve good performance for both models may not be straightforward. 
For separately trained models, each one can have an individual configuration for model architecture and training hyperparameters. Some advanced training techniques more suitable for one type of model can be applied to its training procedure independently. 
For example, discriminative sequence training for the SC model~\cite{Povey2002MinimumPE,Woodland2002LargeSD} and sequence-level objectives for the attention model~\cite{prabhavalkar2018MinimumWordError,cui2018ImprovingAttentionBasedEndtoEnd}.

\presec
\subsection{Decoding}
\postsec
The MAP decoding rule still applies to ISCA to find the 1-best decoding result $\W^{*}$ based on \eqn{isca},
%\begin{align}
%    \W^{*} = & \argmax_{\W} \bigg\{p(\O|\W)P(\W)\sum_{\C\in\mathcal{L}({\W})} P(\C|\O)\bigg\}\nonumber\\
%= & \argmax_{\W}\bigg\{\log p(\O|\W)+\alpha \log P(\W) \nonumber\\
%    & \hspace{5em}+\beta \log\sum_{\C\in\mathcal{L}({\W})} P(\C|\O)\bigg\},
%    \label{eqn:decoding}
%\end{align}
\begin{align}
    &\W^{*} = \argmax_{\W} \bigg\{p(\O|\W)P(\W)\sum_{\C} P(\C|\W,\O)\bigg\}\label{eqn:decoding}\\
    &\approx \argmax_{\W}\bigg\{\ln p(\O|\W)+\alpha\ln P(\W)
 +\beta\ln\sum_{\C} P(\C|\W,\O)\bigg\},\nonumber
\end{align}
where $\alpha$ is the language model scaling factor and $\beta$ is the attention model scaling factor. The use of these scaling factors are necessary for these log-scores to fall within reasonable dynamic ranges. %However, carrying out the exact search according to \eqn{decoding} could be very expensive as the number of $\C$ sequences may be prohibitively large. Therefore, two approximations are used.
To simplify the implementation of ISCA, only a limited number of hypotheses ($n$-best) obtained from the SC-based model are combined with scores from the attention-based model in this paper. %Furthermore, it is worth noting that apart from the traditional frame-synchronised style, the SC decoder can also be an RNN-based decoder, if RNN-transducer can be viewed as an extended SC system.
%Two approximations are taken to simplify the implementation and reduce the computation complexity in decoding. As shown in \eqn{decoding}, the first approximation is that instead of using of all possible word sequences $\W$, only a limited number of hypotheses $\bm{w'}$ are used to compute the attention model score. The second approximation depends on the subword modelling units of the attention model. If the attention subword units are graphemic, no approximation is neede since the mapping $\mathcal{L}(\cdot)$ is one-to-one. However, for phonetic output units, the number of possible sequences can be exponentially large as the utterance gets longer. As an approximation, the most likely sequence $\bm{c'}$ can be obtained by either using Viterbi alignment from a phonetic acoustic model or sampling from all possible pronunciations. 

\presec
\section{Experimental Setup}
\label{sec:setup}
\subsection{Data and Features}
\postsec
In this paper, the individual headset microphone (IHM) from the AMI meeting corpus~\cite{Carletta2005TheAM} is used. The dataset contains around 80 hours of speech for training, and 8 hours for both development (dev) and evaluation (eval). The inputs used are 80-dim filter-bank features at a 10ms frame rate concatenated with 3-dim pitch features.

\presec
\subsection{Model Configurations}
\postsec
The pipeline is based on the ESPnet setup~\cite{watanabe2018espnet,kim2017JointCTCattentionBased}. The default configurations for the acoustic model for the SC model and the encoder of the attention model are both 8-layer bi-directional long short-term memory models (BLSTMs) with a projection layer. The BLSTM has 320 units in each direction and the projection size is also 320. For the SC model, an additional fully connected layer of size 320 is added before softmax output. The attention model uses a location-aware attention mechanism connecting to the decoder with a one-layer 300-unit LSTM.
The n-gram LMs are trained using both AMI and Fisher data. The RNNLM is trained purely based on the text transcriptions of AMI data. The RNNLM has one LSTM layer with 1000 hidden units, whose perplexities on the dev and eval data are 73 and 64 respectively.

\presec
\subsection{Training and Decoding Setups}
\postsec
To train the HMM-based model, the CE objective function is used with alignments produced by a pre-trained DNN-HMM system. Unigram label-smoothing~\cite{Szegedy2016RethinkingTI} is applied before computing the CE loss of the attention model. The numbers of graphemes, monophones and tied triphones are 31, 48 and 4016. The ADADELTA~\cite{Zeiler2012ADADELTAAA} optimiser is used and the batch size is 30 utterances for all models.

For decoding of SC-based models, PyHTK~\cite{Zhang2019PyHTKPL} is used to set up the corresponding HMM structures and the decoding pipeline. HTK~\cite{HTK} tools are used for lattice generation, lattice rescoring with a trigram LM, and $n$-best list generation. %The first pass decoding uses a bigram language model to generate lattices, which are then rescored by a trigram language model in the second pass. 
%A trigram LM is used for decoding SC-based models. 
For decoding of attention models, the width of the beam search is 30.
\presec
\section{Experiments}
\postsec
\label{sec:exp}
In this section, several improvements to the CTC and attention-based model baselines are made to narrow the gap between the traditional DNN-HMM models and the end-to-end models. %Based on the analysis in \sect{scmodel}, the decoding procedure of the HMM-based model is adopted for CTC-based model. 
Next, within the scope of SC-based models, different subword units and objective functions are compared. Under the ISCA framework, both multi-task training and separate training performance are reported. After the exploratory experiments on the AMI dev set, key findings are summarised and the models are tested on the AMI eval set.% in the end. %\tbl{summary}.

\presec
\subsection{Improvements on Multi-task Trained Baseline Systems}
\postsec
In the baseline setup~\cite{kim2017JointCTCattentionBased}, both the CTC model and the attention-based model use the same set of graphemes as their output units and the effective frame rate is 1/4. Frame rate reduction is achieved by skipping some time steps in the LSTM layer. The first LSTM layer has a full frame rate and the next two layers skip one step at every other frame. Statistics on the training data shows that the input/output length ratio is more than four for 95.0\% of all utterances but is more than three for 99.4\% of the data. For phones, 99.9\% of the utterances have an input/output length ratio higher than four. For attention-based models where frame-synchronicity is not required, a higher ratio of frame rate reduction may be appropriate. However, for SC-based models, the frame rate should not be less than 1/3 of the full frame rate.
% \begin{table}[h]
%     \centering
%     \begin{tabular}{lcccc}
%         \toprule
%         \multirow{2}{*}{changes} & standalone && \multicolumn{2}{c}{multi-task}\\
%         \cmidrule{4-5}
%          & Att. && Att. & Att.+CTC \\
%         \midrule
%         baseline & 35.0 && 39.4 & 38.9 \\
%         \midrule
%         +1/3 frame rate & 36.2 && 36.5 & 36.6 \\
%         \;\;+frame shifting & 33.6 && 34.0 & 33.9\\
%         \;\;\;\;+feature concat. & 31.6 && 31.7 &  \\
%         \bottomrule
%     \end{tabular}
%     \caption{Performance improvements over the baseline setup.}
%     \label{tab:improve_train}
% \end{table}
\begin{table}[ht]
    \centering
    \begin{tabular}{l|cc}
        \toprule
         & Att. & Att. + CTC\\
        \midrule
        multi-task baseline & 37.6 & 34.5 \\
        \midrule
        + 1/3 frame rate & 34.9 & 32.2 \\
        %\;\;+ starting frame offsets & 32.7 & 30.1 \\
        \;\;+ training on every frame & 32.7 & 30.1 \\
        % \midrule
        % + 1/3 FR (concat.) & 33.2 & 30.5\\
        \;\;\;\;+ stacking input frames & 30.6 & 28.1 \\
        \bottomrule
    \end{tabular}
    \caption{AMI dev set WERs for graphemic multi-task (CTC \& attention) systems with RNNLM.}
    \label{tab:baseline}
    \vspace{-2em}
\end{table}

In experiments shown in \tbl{baseline}, the frame rate is reduced at the input feature level, which is more similar to the setup from SC-based systems~\cite{Miao2016SimplifyingLS,Povey2016PurelySN}. By sampling one in every three frames, the performance of the model improves despite two-thirds of training data not being used. Offsets on the starting point of the input sequence by one or two frames allow the model to be trained on every frame, which reduces the WER by another 2\% absolute. By stacking two adjacent frames with the current as the input, the WER is further reduced by 2\% since the test-time unused frames are covered~\cite{Pundak2016LowerFR}. Overall, a 6-7\% absolute reduction of WER is observed over the baseline.

\presec
\subsection{Improvements on CTC Models}
\postsec
By treating CTC models as a class of SC-based models, the decoding procedure of the traditional HMM-based systems can be used, where various sources of structured information are incorporated. The baseline uses the prefix search decoding procedure and improvements are made using the lexical-tree-based decoding procedure. 
\begin{table}[ht]
    \centering
    \begin{tabular}{l|c}
        \toprule
         & WER \\
        \midrule
        standalone CTC baseline & 47.6\\
        \midrule
        + graphemic lexicon & 43.9\\
        \;\;+ trigram LM & 38.5\\
        \;\;\;\;+ prior & 33.2\\
        \;\;\;\;\;\;+ multi-task training & 32.1\\
        \bottomrule
    \end{tabular}
    \caption{AMI dev set WER of the standalone CTC model and and several improvements by incorporating structured information during decoding and multi-task training. RNNLM is not used.}
    \vspace{-1em}
    \label{tab:ctc_improve}
\end{table}

As shown in \tbl{ctc_improve}, the addition of a graphemic lexicon that prevents decoding words with incorrect spelling reduces the WER by 2.7\% absolute. However, some words in the lexicon can be decomposed into shorter word-pieces, which happen to also be legal words in the lexicon. The introduction of the trigram LM greatly reduces the fragmentation of words and reduces the WER significantly (5.4\% absolute). One of the many assumptions made by CTC is that all output units have equal prior probabilities. Computed by accumulating the output posteriors from the DNN, the estimated priors are very imbalanced. For example, in the case of 1/3 frame rate, the priors for the blank symbol, the letter `A' and `Z' are $0.43$, $0.03$, $0.0001$ respectively. By using the graphemic unit priors similarly to HMMs as in \eqn{prior}, the WER improves by another 5.3\% absolute. Finally, by training the CTC model and the attention-based model in a multi-task fashion (\cf the same system as the last row in \tbl{baseline}), the CTC WER is further reduced to 32.1\%, whereas the performance of the attention-based model alone is not improved. Narrowing the performance gap between the CTC model and the attention-based model facilitates the use of ISCA framework.%, and the ISCA framework significantly as the decoding procedure is based on the source-channel model.
%Narrowing the performance gap between the CTC model and the attention model helps the ISCA framework significantly as the decoding procedure is based on the source-channel model.

% \subsection{Standalone SC Models}
% \begin{table}[h]
%     \centering
%     \begin{tabular}{c|cc}
%         \toprule
%          & CTC & CE \\
%         \midrule
%         grapheme & 33.2 & --\\
%         monophone & 29.4 & 29.1 \\
%         triphone & 29.6 & 27.0\\
%         \bottomrule
%     \end{tabular}
%     \caption{WERs of standalone SC models with different subword units and loss functions.}
%     \label{tab:unit}
% \end{table}

\presec
\subsection{ISCA for Multi-task Trained Models}
\postsec
Following the multi-task training scheme where the acoustic model (SC-based model) and the encoder (attention-based model) are shared, the following experiments vary the modelling units and objective functions of the SC-based model, while keeping those of the attention-based model unchanged. In the following ISCA results, 20-best hypotheses from the SC-based models are used as an approximation to \eqn{decoding}. This simplified ISCA ranks the 20-best list by optimally interpolating the acoustic scores, trigram LM scores, the attention-based model scores, and optionally the RNNLM scores. A derivative-free stochastic global search algorithm, named covariance matrix adaptation evolution strategy (CMA-ES)~\cite{Hansen2001CompletelyDS}, is used to optimise the interpolation weights.
\begin{table}[ht]
    \centering
    \begin{tabular}{cc|ccl}
        \toprule
        subword units & loss & SC & Att. & ISCA (+RNNLM) \\
        \midrule
        grapheme & CTC & 32.1 & 31.7 & \;28.6\;\;(28.4)\\
        \midrule
        \multirow{2}{*}{monophone} & CTC & 28.8 & 31.6 & \;26.7\;\;(26.3)\\
         & CE & 28.8 & 31.8 & \;26.5\;\;(26.0)\\
        \midrule
        \multirow{2}{*}{triphone\tablefootnote{A decision tree is used to tie 93k triphone models (or states) to 4k.}} & CTC & 28.3 & 31.4 & \;26.2\;\;(25.6)\\
         & CE & 26.6\tablefootnote{An HTK acoustic model can yield similar WERs with only a fifth of the parameters due to better data shuffling, larger batch size \etc. For results to be comparable, all models in this paper are trained using ESPnet.} & 31.6 & \;25.3\;\;(24.7)\\
        \bottomrule
    \end{tabular}
    \caption{AMI dev set WERs of multi-task trained systems. CE systems use single-state HMMs. A trigram LM is used for SC-based model decoding. RNNLM is not used except for results in brackets.}
    \label{tab:multitask}
    \vspace{-1em}
\end{table}

The model in the first row in \tbl{multitask} corresponds to the model of the last rows in both \twotbl{baseline}{ctc_improve}\footnote{It is a multi-task trained model with graphemic CTC and attention. Systems in \tbl{baseline} have an RNNLM while \twotbl{ctc_improve}{multitask} do not. \twotbl{baseline}{multitask} shows comparable results for joint decoding (28.1) and ISCA (28.4).}. For this multi-task trained grapheme/CTC and attention-based model, the performance of the ISCA framework is similar to the joint decoding result in \tbl{baseline}. %However, the performance of the grapheme/CTC model is relatively poor.

Given that the joint decoding method only applies to a CTC model and an attention-based model with the same graphemic units, one of the major advantages of the ISCA framework is that the SC model can have any subword units and any loss functions. The next two rows change the modelling units of SC-based systems to monophones, and the WER of the SC-based system reduces by 10.3\% relative to the grapheme/CTC model. This is mainly due to the orthographic irregularity in English where phonetic-based units reduce the difficulty of modelling significantly. The mapping between pronunciation and text is achieved by the lexicon embedded in the decoding graph. For these two monophone SC models, improved CTC and CE models have similar performance. However, this observation does not hold for triphone systems. This is possibly because triphone systems are more sensitive to the quality of the training alignments since there are many more confusing triphone units than the monophone units. On the scale of AMI data set, the CTC system requires more training data to learn the triphone alignments properly by itself, whereas the CE system shows its advantage whose alignments were produced by a pre-trained high-performance system.
%The CTC model may be more sensitive to the quality of the decision tree as a badly clustered triphone may generate many possible confusing alignments using the CTC loss function. In contrast, for the CE criterion, some poorly clustered triphone states are equivalent to introducing some noisy labels. By comparing the WERs of various SC-based models in \tbl{multitask}, the triphone/CE model outperforms all others with different subword units and/or loss functions, despite that a forced alignment from an existing system is required. 
%However, this oberservation does not hold for SC-based triphone systems. The CTC model may be more sensitive to the quality of the decision tree as a badly clustered triphone may generate many possible confusing alignments using the CTC loss function. In contrast, for the CE criterion, some poorly clustered triphone states are equivalent to introducing some noisy labels. By comparing the WERs of various SC-based models in \tbl{multitask}, the triphone/CE model outperforms all others with different subword units and/or loss functions, despite that a forced alignment from an existing system is required. 
%It is also worth noting that the triphone/CE performance is not comparable to the standard settings in the traditional hybrid model training pipeline, where frame-level shuffling with a large batch size is not possible under the multi-task training framework. 

Further experiments show that multi-task trained SC-based models perform marginally better than their standalone counterparts. However, the expected benefits of multi-task training are not observed for the attention-based models. The last column in \tbl{multitask} shows that the ISCA approach yields consistent reduction in WER while the SC-based model improves, with or without an external RNNLM. However, as the performance gap between the SC-based model and the attention-based model widens, the relative improvement \wrt the SC-based model shrinks. In the extreme case where the triphone/CE model outperforms the attention-based model by 15.8\% relative, ISCA with 20-best rescoring can still improve the WER of the SC-based model by 4.8\% relative.

Since the ISCA framework integrates two models at the word-level, attempts have also been made to change the modelling units of the attention-based model. Similar to the findings in~\cite{sainath2018NoNeedLexicon,zhou2018SyllableBasedSequencetoSequenceSpeech}, using monophone for the attention-based model is not as helpful as graphemes for ISCA, which shows that the complementarity of graphemic and phonetic models is essential. Since the neural decoder directly conditions on the previous output as in \eqn{neuraldecoder}, using context-dependent units for the attention-based model yields essentially no improvement compared to their context-independent counterparts as expected. For the attention-based models, modelling context-dependent phones is even a harder task than context-independent ones as the model also needs to learn the tying results found by phonetic decision trees. Another issue of using phonetic units for attention-based models is that, as a result of multiple pronunciations, there may be an exponential number of possible sequences $\C$ in \eqn{decoding} when the utterance gets longer.

% \begin{table}[h]
%     \centering
%     \begin{tabular}{c|ccc}
%         \toprule
%         subword & SC~(tg) & ISCA~(tg) & ISCA~(RNNLM) \\
%         \midrule
%         grapheme & 26.6 & 25.3 & 24.7\\
%         mono-PEG & 26.5 & 25.4 & 24.7\\
%         tri-PEG & 26.7 & 26.1 & 25.4\\
%         monophone & 26.5 & 26.1 & 25.2\\
%         triphone & 26.7 & 26.5 & 25.5\\
%         \bottomrule
%     \end{tabular}
%     \caption{ISCA performance of triphone/CE SC model and attention model with different subword units.}
%     \label{tab:seq2sequnits}
% \end{table}

\presec
\subsection{ISCA for Models Trained Separately}
\postsec
Since the triphone/CE model outperforms the attention-based model significantly under the multi-task training setup and only 20-best hypotheses are used for ISCA, two more questions remain. First, how ISCA performs when the attention-based model improves. Second, if the first approximation in \eqn{decoding} becomes more accurate, \ie the $n$-best list becomes longer, how much more improvement ISCA can make. In order to answer these questions, three separate models are trained whose configurations are listed in \tbl{sepmodels}.
% (1) a triphone/CE model which is of the same size as the previous triphone/CE branch in \tbl{multitask} (WER=27.0); (2) a small attention model which is of the same size as the previous attention branch in \tbl{multitask} (WER=31.6 without RNNLM); (3) a large attention model with 4 BLSTM layers of size 1024, whose number of parameter is more than fivefold than the small attention model (WER=26.3).
\begin{table}[ht]
    \centering
    \begin{tabular}{ll|c}
        \toprule
        model & BLSTM config. & WER (+RNNLM) \\
        \midrule
        SC(triphone/CE) & 8-layer, 320 units & 27.0\;\;(25.8) \\
        Att.(small) & 8-layer, 320 units & 31.6\;\;(30.2) \\
        Att.(large) & 4-layer, 1024 units & 26.3\;\;(25.8) \\
        \bottomrule
    \end{tabular}
    \caption{Configs. and WERs of three separately trained models.}
    \label{tab:sepmodels}
    \vspace{-2em}
\end{table}
\begin{figure}[ht]
    \centering
    \includegraphics[width=0.98\linewidth]{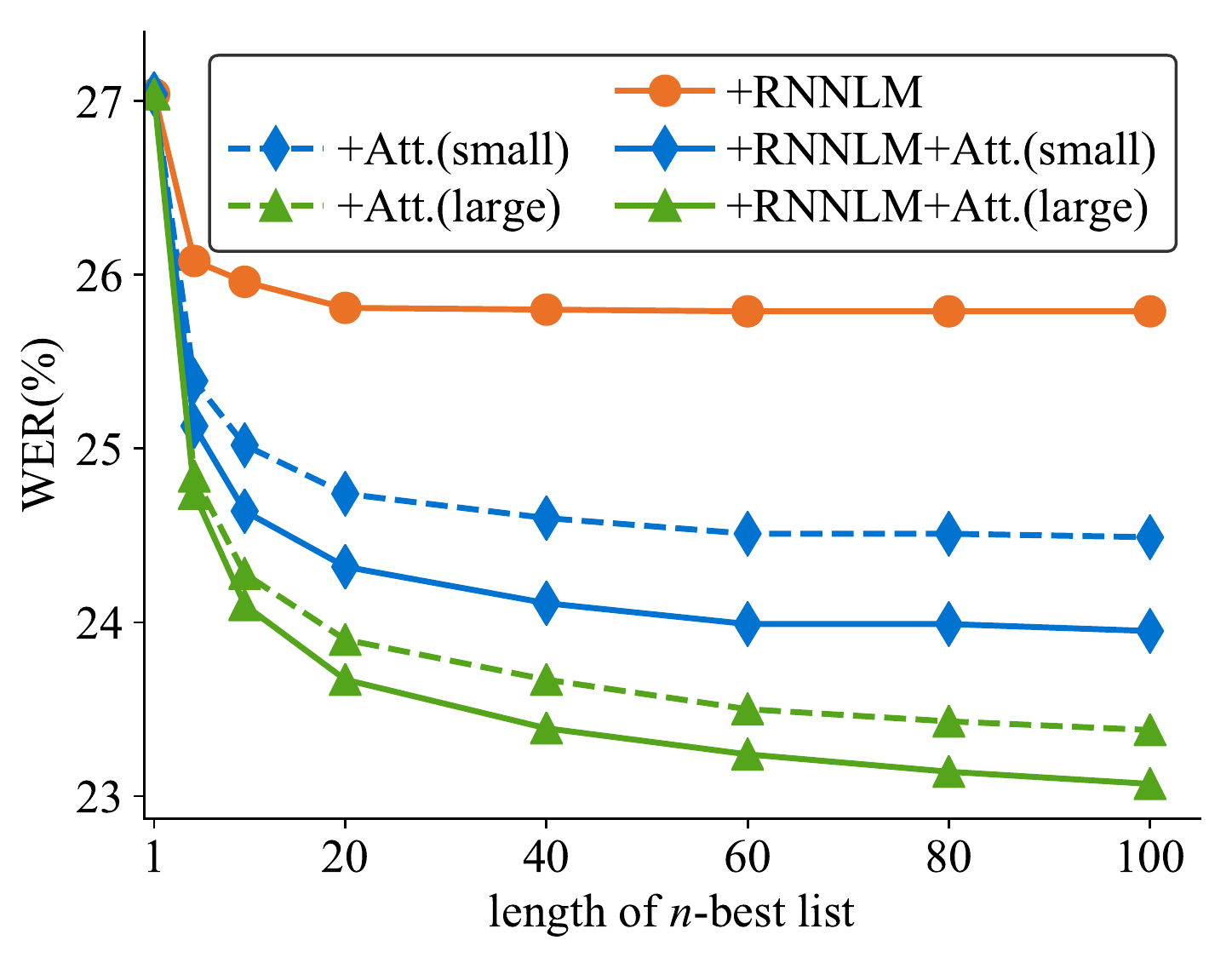}
    \vspace{-1em}
    \caption{ISCA between a separately trained triphone/CE model and attention-based models of different sizes.}
    \label{fig:nbest}
    \vspace{-1em}
\end{figure}

As shown in \fig{nbest}, the reduction in WER by an external RNNLM stagnates after an $n$-best list of length 20 is used. However, the WER of ISCA continues to drop, especially for the large attention-based model where two systems have similar WERs. Though the attention-based models the acoustic and language model jointly, an RNNLM trained using the AMI transcription only is still useful. The improvement by the RNNLM is expected to be greater when additional text corpora in similar domains are used for training. For the 100-best list, small and large attention-based models with the RNNLM reduces the WER by 7.1\% and 10.5\% relative to the triphone/CE model rescored by the RNNLM.

\presec\presec
\subsection{Summary and Discussions}
\postsec
Key results are summarised in \tbl{summary}. The total number of model parameters and performance on the AMI eval set are also reported for the following models and setups.
\begin{table}[ht]
    \centering
    \begin{tabular}{ll|r|cc}
        \toprule
        & & \#params. & dev & eval\\
        \midrule
        \multirow{2}{*}{multi-task} & baseline~\cite{kim2017JointCTCattentionBased} & 15.6M & 34.5 & 37.4\\
         & improved baseline & 16.1M & 28.1 & 29.2 \\
        \midrule
        \multirow{3}{*}{separate} & SC (triphone/CE)\tablefootnote{The best performing system reported on AMI dev and eval sets is SC-based with WER about 19\%, where speaker adaptive training, sequence training, speed perturbation and many other techniques have been used \cite{Cheng2017AnEO}.} & 16.0M & 25.8 & 26.8\\
        %\multirow{3}{*}{separate} & SC (triphone/CE)\tablefootnote{The state-of-the-art SC-based system gave WERs being around 19\%, where lattice-free discriminative sequence training \textit{etc}. were used.} & 16.0M & 25.8 & 26.8\\
         & Att.(small) & 15.9M & 30.2 & 31.0\\
         & Att.(large) & 84.0M & 25.8 & 25.9\\
        \midrule
        \multirow{3}{*}{ISCA} & multi-task  & 17.3M & 24.4 & 25.4\\
        % \multirow{3}{*}{ISCA} & multi-task training & 17.3M & 24.7 & 25.8\\
        % & SC + Att.(small) & 24.3 & 25.1\\
        % & SC + Att.(large) & 23.7 & 24.6\\
        % \midrule
         & SC + Att.(small) & 31.9M & 24.0 & 24.5 \\
         & SC + Att.(large) & 100.0M & \bf{23.1} & \bf{23.8}\\
        \bottomrule
    \end{tabular}
    \caption{The number of parameters and WERs of some key models on both dev and eval sets, with RNNLM included. All ISCA systems use a 100-best list produced by the triphone/CE SC-based model.}%All ISCA systems used 100-best from the triphone/CE model.}
    \vspace{-1em}
    \label{tab:summary}
\end{table}
By changing the encoder frame rate and arrangement of input acoustic features, the joint decoding WER of the multi-task trained CTC and the attention-based model drops from 37.4\% to 29.2\%. For standalone SC-based models, triphone/CE models consistently outperform other types of SC-based models. The standalone attention-based model requires five-fold more model parameters to reach competitive performance against the triphone/CE model. Although having more parameters combined, separately trained models outperforms the multi-task trained model with a similar model size per branch. And the flexibility of individual optimisation leads to greater improvement when the performance of the two models are closer. Comparing to the improved baseline, ISCA reduces the WER by 13\% relative for the multi-task trained system. For separately trained models, ISCA achieves relative WER reductions of 8.6\% and 21\% \wrt the triphone/CE model and the small attention-based model; 11.2\% and 8.1\% \wrt the triphone/CE model and the large attention-based model. Further improvement is expected if longer $n$-best lists are used or using ISCA for lattice rescoring~\cite{liu2016two}, and if speaker adaptation is applied to both SC-based systems~\cite{gales1998maximum,saon2013speaker,li2018SpeakerAdaptationEndtoEnd} and attention-based systems~\cite{delcroix2018AuxiliaryFeatureBased}.
\presec
\section{Conclusions}
\postsec
\label{sec:conclusion}
In this paper, we have proposed a flexible framework called ISCA that combines a source-channel model system with an attention-based system viewed as a special language model conditioned on acoustic information. Since the two highly complementary models are integrated at the word-level, each one can be trained independently with different objective functions and/or lexicons, or they can be optimised in a multi-task training fashion. As the performance gap between attention-based systems and the traditional hybrid systems narrows, ISCA is expected to yield the state-of-the-art performance when combining the best system from both sides, while improving the robustness of the overall system.

\vfill
\pagebreak

\presec
\section{References}
\vspace{-0.5em}
\begingroup
\renewcommand{\section}[2]{}
\bibliographystyle{IEEEbib}
\bibliography{refs}
\endgroup

\end{document}